\documentclass[pra,twocolumn,superscriptaddress,amsmath,amssymb,preprintnumbers,showpacs,letterpaper,aps,10pt,nofootinbib]{revtex4-1}
\providecommand{\bra}[1]{\langle #1 |}
\providecommand{\ket}[1]{| #1 \rangle}

\def\vec#1{\text{\bfseries#1}}
\usepackage{graphicx,color,bm,amsmath,amssymb}
\DeclareMathOperator{\tr}{tr}

\DeclareMathOperator{\rank}{rank}

\DeclareMathOperator{\sgn}{sgn}

\begin{document}

\title{Entanglement in bipartite pure states of an interacting boson gas obtained by local projective measurements}
\author{Francis N.~C.~Paraan}
\email[Electronic address: ]{\url{fparaan@max2.physics.sunysb.edu}}	
\affiliation{C. N. Yang Institute for Theoretical Physics, State University of New York at Stony Brook, NY 11794-3840, USA}

\author{Javier Molina-Vilaplana}
\email[Electronic address: ]{\url{javi.molina@upct.es}}
\affiliation{Department of Systems Engineering and Automation, Technical University of Cartagena, 30202 Cartagena, Spain}

\author{Vladimir E.~Korepin}
\email[Electronic address: ]{\url{korepin@max2.physics.sunysb.edu}}	
\affiliation{C. N. Yang Institute for Theoretical Physics, State University of New York at Stony Brook, NY 11794-3840, USA}

\author{Sougato Bose}
\email[Electronic address: ]{\url{sougato@theory.phys.ucl.ac.uk}}
\affiliation{Department of Physics and Astronomy, University College London, London WC1E 6BT, UK}
\date{August 8, 2011}
\preprint{YITP-SB-11-14}

\begin{abstract}
We quantify the extractable entanglement of excited states of a Lieb-Liniger gas that are obtained from coarse-grained measurements on the ground state in which the boson number in one of two complementary contiguous partitions of the gas is determined. Numerically exact results obtained from the coordinate Bethe ansatz show that the von Neumann entropy of the resulting bipartite pure state increases monotonically with the strength of repulsive interactions and saturates to the impenetrable boson limiting value. We also present evidence indicating that the largest amount of entanglement can be extracted from the most probable projected state having half the number of bosons in a given partition. Our study points to a fundamental difference between the nature of the entanglement in free-bosonic and free-fermionic systems, with the entanglement in the former being zero after projection, while that in the latter (corresponding to the impenetrable boson limit) being non-zero.
\end{abstract}

\maketitle
\hbadness=10000
\section{Introduction}
In the quantum information and communication community, studies of the properties of quantum entanglement and entanglement measures are mainly motivated by the idea that entanglement is a vital resource for processing quantum information. Of particular current interest is the quantification of the entanglement present in many-body quantum systems \cite{[{For reviews see }] eisert,*amico} for which experimental measurement protocols have been recently proposed \cite{cramer}. In this paper we will study the entanglement that can be extracted from bipartite projections of the ground state of the integrable repulsive Lieb-Liniger model on a ring \cite{lieb1,lieb2}. The corresponding hamiltonian describes a continuous one-dimensional gas of bosons with short-range interactions modeled by Dirac delta-functions. Though these one-dimensional systems have long been studied in atomic waveguides with tight axial confinement \cite{gorlitz,paredes,kinoshita}, the ring-shaped traps needed to reproduce periodic boundary conditions have only been recently developed \cite{morizot}. 

The main objective of our work here is to study the effects of the strength of interactions on the entanglement obtained from projections of the ground state
of this continuous many-body system---an aim that is facilitated by the fact that the Lieb-Liniger hamiltonian is exactly solvable by the Bethe ansatz. In fact, the entanglement entropy in other continuous integrable systems, for example the Calogero-Sutherland model \cite{haque,katsura} and the anyonic Lieb-Liniger model \cite{haque,guo,santachiara}, has already been studied under the framework of particle partitioning, but as we explain in the following paragraphs our extraction procedure involves a spatial partitioning of subsystems. More recently, the von Neumann and R\'enyi block entropies in the ground state of spatially continuous free-fermionic models \cite{calabrese2,*calabrese3} were obtained from conformal field theory arguments as the appropriate limit of the corresponding lattice model results \cite{fagotti,calabrese1}.

Although there is as yet no conventional approach to quantifying the amount of entanglement in multi-particle assemblies of indistinguishable particles \cite{sasaki,dowling,wiseman}, the type of projections we consider here allows us to give a definite measure of entanglement in the resulting projected states via the von Neumann entropy. In such indistinguishable particle systems, entanglement between two spatial partitions of the system may both arise from the correlation in occupancies of the partitions, as well as the correlation between distributions of the particles in the two partitions. For example, occupancies of two boxes, by a total number of two indistinguishable bosons can be of the form $\ket{2,0}+\ket{1,1}+\ket{0,2}$,\footnote{The state ket $\ket{m,n}$ denotes the state having $m$ particles in one box and $n$ in the other.} and there is entanglement in this state by virtue of the uncertainty of the number of particles in each box. If one can somehow eliminate this form of (somewhat trivial) entanglement, then does one still retain any entanglement between partitions in a system of indistinguishable particles? We will show below that the answer to this question is ``no'' for free bosons, while it is ``yes'' for free fermions, providing a striking qualitative difference in the cause of the entanglement in these systems. 

In this paper, we are interested in quantifying the entanglement in projections of the Lieb-Linger ground state that are obtained after coarse-grained measurements reveal the number of particles in one of two contiguous partitions $A$ and $B$ that divide the gas into two half-rings. Thus, the projected pure state is spatially partitioned (bipartite) and one can quantify the resulting entanglement by the von Neumann entropy. A similar procedure has been used to quantify the entanglement extractable from stationary and non-stationary states of impenetrable boson gases \cite{molina1}, supersinglet states, and several spin chains \cite{molina2}. In the spirit of these latter works we study here the notion of a projectively extractable pure state entanglement $\mathcal{E}_{P\negthinspace P}$ as defined by
\begin{equation}\label{epp}
	\mathcal{E}_{P\negthinspace P} \equiv \max_{k}\,\{p(k)S_{A}(k)\} \equiv \max_{k}\,\{\mathcal{E}_k\},
\end{equation}
where $p(k)$ is the probability of projecting the ground state into a state $\chi_{AB}(k)$ having the fraction $k/N$ of particles in region $A$ and $S_A(k)\equiv -\tr[\rho_A(k)\log_2\rho_A(k)]$ is the von Neumann entropy of the reduced density matrix $\rho_A (k)\equiv \tr_B\ket{\chi_{AB}(k)}\bra{\chi_{AB}(k)}$. That is, if $\Pi_k$ is a projector onto the state subspace having $k$ particles in partition $A$, then the projected ket we are interested in may be expressed as $\ket{\chi_{AB}(k)} = \Pi_k\ket{\chi}$ in terms of the ground state ket $\ket{\chi}$. The entanglement measure above \eqref{epp} gives the maximum weighted entanglement $\mathcal{E}_k \equiv p(k)S_{A}(k)$ over all possible projection outcomes and consequently captures the probabilistic nature of the preparatory measurements.

Our analysis begins in Sect.~\ref{bethe} with a brief introduction to the coordinate Bethe ansatz that forms the basis of our exact computations. The probability of obtaining each projection outcome is calculated in Sect.~\ref{probsect}, while the von Neumann entropy of the projected states and their corresponding weighted entanglement are given in Sect.~\ref{vnesect}. On the basis of these results we argue that the projectively extractable pure state entanglement $\mathcal{E}_{P\negthinspace P}$ is equal to the weighted entanglement $\mathcal{E}_{N/2}$ of the balanced case in which exactly half of the bosons are present in both partitions (the total number $N$ is given to be even). Our main conclusion is that the extractable entanglement $\mathcal{E}_{P\negthinspace P}$ increases monotonically with interaction strength and saturates to its impenetrable boson value. We conclude with a summary and some remarks in Sect.~\ref{conclusions}.

\section{Bethe ansatz}\label{bethe}
We briefly review here some of the well-established properties of the Bethe ansatz solution for the ground state of the Lieb-Liniger model \cite{lieb1,korepin1}. Written in dimensionless form, in which lengths are measured in units of the ring circumference $L$ and energy in natural units $\hbar^2/2mL^2$, the Schr\"odinger equation for $N$ delta-interacting bosons is
\begin{equation}\label{hamil}
	\mathcal{H}\chi(\vec{x}) = \biggl[-\sum_{j=1}^N\frac{\partial^2}{\partial x_j^2} + 2c\negthickspace\sum_{1\le k<j\le N}\negthickspace\delta(x_j-x_k)\biggr]\chi(\vec{x}).
\end{equation}
The dimensionless interaction constant is taken to be non-negative $c\ge0$ so that the gas does not collapse. The unique periodic ground state eigensolution of this hamiltonian is given by the normalized coordinate Bethe ansatz 
\begin{align}
	\chi(\vec{x}) &= \frac{1}{\mathcal{N}}\sum_{\{\mathcal{P}\}}^{N!} (-1)^{[\mathcal{P}]} F_\mathcal{P}(\vec{x})e^{i{\bm\lambda}_\mathcal{P}\cdot\vec{x}},\\ 
	F_\mathcal{P}(\vec{x})&\equiv\prod_{1\le k<j\le N}\frac{\lambda_{\mathcal{P}j}-\lambda_{\mathcal{P}k}-ic\sgn(x_j-x_k)}{\bigl\{N!\prod_{k<j}\bigl[(\lambda_j-\lambda_k)^2+c^2\bigr]\bigr\}^{1/2}},
\end{align}
where the momentum vector ${\bm\lambda}_\mathcal{P}$ has $N$ components $\lambda_{\mathcal{P}j}$ that form a permutation $\mathcal{P}$ of the solutions $\lambda_j$ of the Bethe equations
\begin{equation}\label{betheeqn}
	e^{i\lambda_j} = -\prod_{k=1}^N\frac{\lambda_j-\lambda_k+ic}{\lambda_j-\lambda_k-ic},
\end{equation}
which satisfy $\sum_i\lambda_i = 0$. The quantity $(-1)^{[\mathcal{P}]}$ is the signature of the permutation $\mathcal{P}$ and the normalization factor $\mathcal{N}$ may be obtained by, for example, the quantum inverse scattering method \cite{korepin2}. Explicitly, the absolute square $\left|\mathcal{N}\right|^2$ is the determinant of the second derivatives of the Yang action $S$ evaluated at the solutions of the Bethe equations \eqref{betheeqn}:
\begin{align}
	\left|\mathcal{N}\right|^2 &= \det\biggl(\frac{\partial^2 S}{\partial\lambda_j\partial\lambda_k} \biggr)  \nonumber\\
	&= \det\biggl(\delta_{jk} + \sum_{l=1}^{N}\frac{2c\delta_{jk}}{(\lambda_k-\lambda_l)^2+c^2} \nonumber \\
	&\qquad\qquad\qquad\qquad\qquad-\frac{2c}{(\lambda_j-\lambda_k)^2+c^2}\biggr).
\end{align}

\section{Projection probabilities}\label{probsect}

\begin{figure}[tb]
	\centering
		(a)\includegraphics[width=0.85\linewidth]{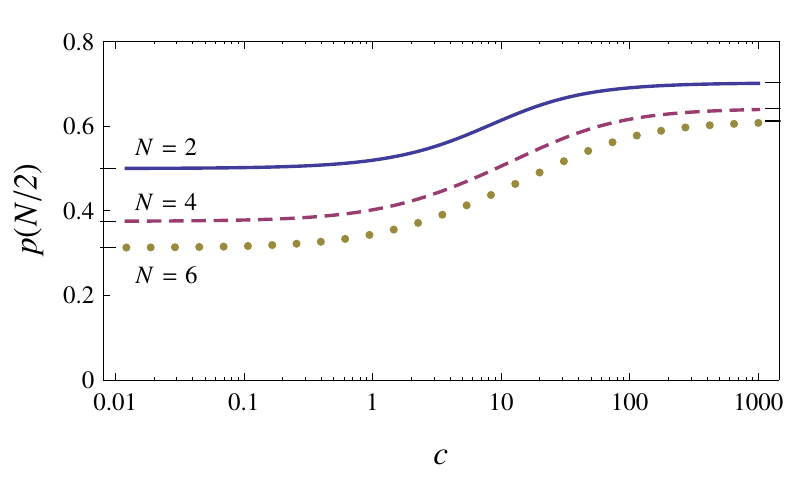}\\(b)\includegraphics[width=0.85\linewidth]{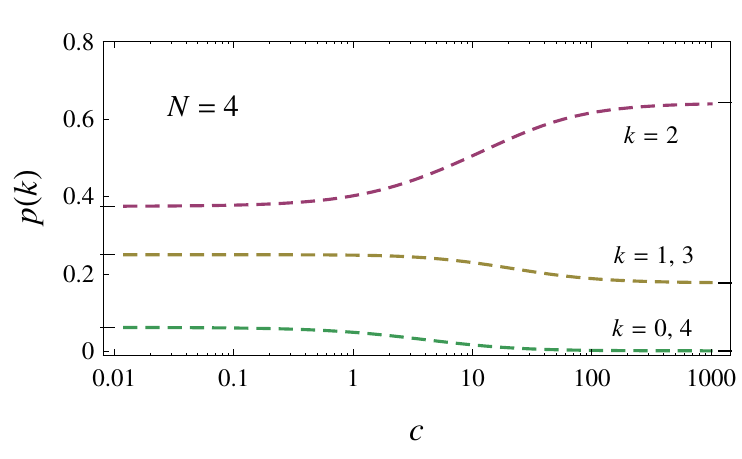}
	\caption{(Color online) (a) The probability of finding an equal number of particles in each half-ring $A$ and $B$ in the Lieb-Liniger ground state increases monotonically with repulsion strength and decreases with particle number. (b) With $N=4$ bosons the balanced case $k=2$ is the most probable situation for all repulsion strengths. Black tick marks denote the free and impenetrable limit values.}\label{probab}
\end{figure}

In this section we are interested in calculating the probability $p(k)$ of successfully projecting the ground state $\chi$ into the pure state $\chi_{AB}(k)$ having $k$ bosons in partition $A$ and $N-k$ bosons in partition $B$. For convenience let us define the vectors $\vec{x}_A \equiv (x_1,\dotsc,x_{k})^\text{\sffamily{T}}$ and $\vec{x}_B \equiv (x_{k+1},\dotsc,x_N)^\text{\sffamily{T}}$. Since the wavefunction $\chi(\vec{x})$ is manifestly invariant over any permutation of coordinate indices, we can express the success probability as
\begin{equation}
	p(k) = \binom{N}{k}\int_A\int_B |{\chi(\vec{x})}|^2\,d\vec{x}_B d\vec{x}_A.
\end{equation}
Also, because of the translational invariance of the ground state wavefunction we may choose the partitions to be $A=\{x|0\le x\le\tfrac{1}{2}\}$ and $B=\{x|\tfrac{1}{2}\le x \le1\}$ without loss of generality. 

\begin{figure}[tb]
	\centering
		\includegraphics[width=0.85\linewidth]{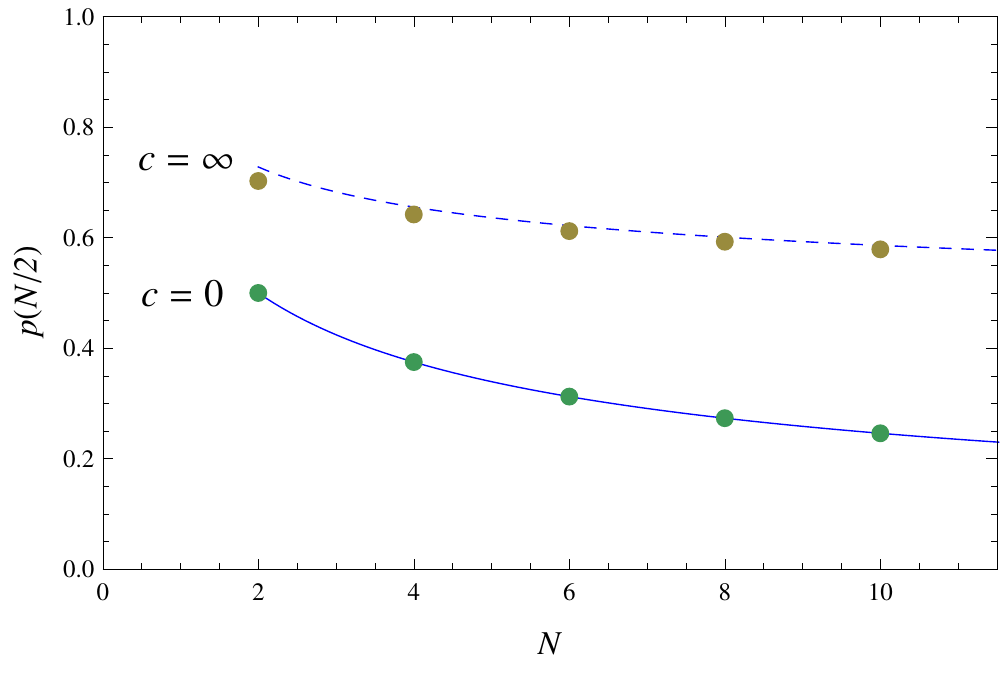}
	\caption{(Color online) Balanced projection probabilities for the free boson ($c=0$) and impenetrable boson ($c\to\infty$) cases. Our numerical values (dots) are compared to the exact result for free bosons (solid line) and the asymptotic result Eq.~\eqref{probcinfinity} for impenetrable bosons (dashed line).\label{probasymfig}}
\end{figure}

Fig.~\ref{probab}(a) shows the probability of projection onto the balanced $k=N/2$ bipartite states for two, four, and six bosons at arbitrary repulsion strengths. For free bosons ($c=0$) this probability is equal to $N!(N/2)!^{-2}2^{-N}$ (Fig.~\ref{probasymfig}) and vanishes as $\sim\negmedspace\sqrt{2/(\pi N)}$ in the thermodynamic limit $N\to\infty$. This is the expected result because each independent particle may be found in either half-ring with equal probability. 
 
As the interparticle repulsion is turned on, however, correlations arise between the positions of the bosons and the corresponding success probability deviates from the free boson value. These quantum correlations give rise to fluctuations in the number of particles in each partition that result in the probability $p(N/2)$ increasing with repulsion strength $c$. In the limiting case $c\to\infty$ of impenetrable bosons $\bm($also known as the Tonks-Girardeau (TG) limit$\bm)$ this probability reduces to the analogous projection probability in a free spinless fermion gas \cite{girardeau}. The characteristic function of this probability distribution of particle numbers in finite regions of an infinite line \cite{aristov,abanov} and a ring \cite{ovchinnikov} are known. We find that a Gaussian approximation (details are given in the Appendix) to the probability distribution $p(k)$ asymptotically yields a balanced projection probability of 
\begin{equation}\label{probcinfinity}
	p(N/2) \sim \sqrt{\frac{\pi}{2\ln(2Ne^{\gamma+1})} }, \quad N\gg 1.
\end{equation}
This probability decays to zero logarithmically for large $N\to\infty$ (Fig.~\ref{probasymfig}). It is clear that this probability distribution is symmetric about $N/2$ because $p(k) = p(N-k)$, and we further expect it to be unimodal.

For general $k$ the projection probabilities in the free boson limit are equal to 
\begin{equation}
	p(k)= \binom{N}{k}\frac{1}{2^N}, \quad c=0,
\end{equation}
while in the TG limit we expect this distribution to be peaked and centered about $k=N/2$ for a fixed large $N$. Therefore, at these two extreme limits the most probable result of the projective measurement is the balanced bipartite state $\chi_{AB}(N/2)$. Since there are no critical values of $c\in(0,\infty)$, we do not expect this trend to change for arbitrary finite repulsion strengths and we argue that the balanced projection is the most probable outcome for all non-negative values of $c$. We confirm this conjecture for $N=4$ bosons and show in Fig.~\ref{probab}(b) all possible projection probabilities for arbitrary repulsion strengths.

\section{von Neumann entropy and extractable entanglement}\label{vnesect}
We now calculate the entanglement between the partitions $A$ and $B$ by calculating the von Neumann entropy $S_A(k)$ of region $A$. The joint density matrix of the projected pure state is $\rho_{AB} = \ket{\chi_{AB}}\bra{\chi_{AB}}$ and the relevant reduced density matrix is $\rho_A = \tr_B\rho_{AB}$. In the coordinate basis where $\ket{\vec{x}} \equiv \psi^\dag(x_N)\dotsb\psi^\dag(x_1)\ket{0}$ with $\psi^\dag(x)$ a bosonic field creation operator and $\ket{0}$ the vacuum state ket, the joint density matrix may be written as 
\begin{align}\label{joint}
	\rho_{AB}(k) &= \binom{N}{k}\frac{1}{p(k)|\mathcal{N}|^2}\sum_{\{\mathcal{P\}}\{\mathcal{Q\}}}(-1)^{[\mathcal{P}]+[\mathcal{Q}]} \nonumber  \\
	&\quad\times\int_A\int_A\int_B\int_B F_\mathcal{P}(\vec{x})\bar{F}_\mathcal{Q}(\vec{x}')e^{i({{\bm\lambda}_\mathcal{P}\cdot\vec{x}}-{{\bm\lambda}_\mathcal{Q}\cdot\vec{x}'})} \nonumber \\
	&\quad \quad\times \ket{\vec{x}}\bra{\vec{x}'}\,d\vec{x}_Bd\vec{x}_B'd\vec{x}_Ad\vec{x}_A'.
\end{align}
The signum functions in the amplitudes $F_\mathcal{P}(\vec{x})$ evaluate trivially for each case $\sgn(x_{Bj}-x_{Ak}) =1$ so that it factorizes according to $F_\mathcal{P}(\vec{x}) \equiv F_{\mathcal{P}\negthinspace A}(\vec{x}_A)F_{\mathcal{P}\negthinspace B}(\vec{x}_B)$. Thus, the integrals over region $B$ may be evaluated independently and tracing away the degrees of freedom in $B$ gives the reduced density matrix
\begin{multline}\label{reduced}
	\rho_{A}(k) = \sum_{\{\mathcal{P\}}\{\mathcal{Q}\}}
	\int_A\int_A G_{\mathcal{P}}(\vec{x}_A)\bar{G}_{\mathcal{Q}}(\vec{x}_A')\\\times e^{i({{\bm\lambda}_{\mathcal{P}\negthinspace A}\cdot\vec{x}_A}-{{\bm\lambda}_{\mathcal{Q}A}\cdot\vec{x}_A'})}\ket{\vec{x}_A}\bra{\vec{x}'_A}\,d\vec{x}_Ad\vec{x}'_A,
\end{multline}
where the function $G_{\mathcal{P}}(\vec{x}_A)$ is
\begin{multline}
	G_{\mathcal{P}}(\vec{x}_A) =\sqrt{\binom{N}{k}\frac{1}{p(k)|\mathcal{N}|^2}}\,(-1)^{[\mathcal{P}]}F_{\mathcal{P}\negthinspace A}(\vec{x}_A) \\
	\times \int_B F_{\mathcal{P}\negthinspace B}(\vec{x}_B)e^{i{\bm\lambda}_{\mathcal{P}\negthinspace B}\cdot\vec{x}_B}d\vec{x}_B.
\end{multline}
and we have introduced the permuted momentum vectors ${\bm \lambda}_{\mathcal{P}\negthinspace A} \equiv (\lambda_{\mathcal{P}1},\dotsc,\lambda_{\mathcal{P}k})^\text{\sffamily{T}}$ and ${\bm \lambda}_{\mathcal{P}\negthinspace B} \equiv (\lambda_{\mathcal{P}k+1},\dotsc,\lambda_{\mathcal{P}N})^\text{\sffamily{T}}$. The eigenvalues $a_i$ of the reduced density matrix $\rho_A$ may be obtained by diagonalizing the associated homogeneous Fredholm integral equation
\begin{equation}\label{eigen}
	\int_A	 K(\vec{x}_A,\vec{x}_A')\phi_i(\vec{x}_A')\,d\vec{x}_A' = a_i\phi_i(\vec{x}_A),
\end{equation}
which has a degenerate kernel
\begin{equation}
	K(\vec{x}_A,\vec{x}_A') \equiv\negthickspace \sum_{\{\mathcal{P\}}\{\mathcal{Q}\}}\negthickspace G_{\mathcal{P}}(\vec{x}_A)e^{i{{\bm\lambda}_{\mathcal{P}\negthinspace A}\cdot\vec{x}_A}}\bar{G}_{\mathcal{Q}}(\vec{x}_A')e^{-i{{\bm\lambda}_{\mathcal{Q}A}\cdot\vec{x}_A'}}.
\end{equation}

\begin{figure}[tb]
	\centering
		\includegraphics[width=0.9\linewidth]{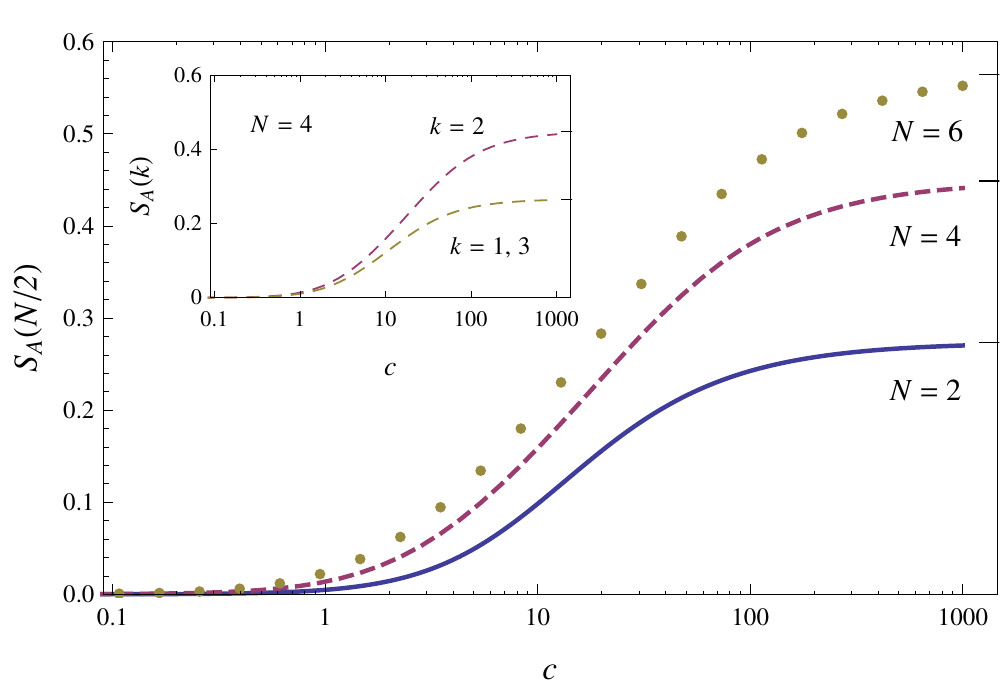}
	\caption{(Color online) The von Neumann entropy between half-rings in the projected balanced pure state increases monotonically with the strength of repulsion. For the case of four bosons the balanced projection $k=N/2$ is more entangled than the unbalanced cases for all $c$ (inset). Black tick marks denote the corresponding values in the impenetrable boson limit.\label{entanglement}}
\end{figure}

Since free bosons condense in the ground state, the kernel reduces to a constant when there are no interactions and the only eigenvalue of the resulting reduced density matrix is unity. Thus, there is no entanglement when $c=0$. For any non-zero contact repulsion between $N$ bosons, however, there are $\binom{N}{k}$ distinct ways of choosing the components of the vectors $\bm\lambda_{\mathcal{P}\negthinspace A}$ and $\bm\lambda_{\mathcal{Q}A}$. From the observation that the plane waves $e^{i\lambda_{\mathcal{P}i}x_i}$ are linearly independent and square-integrable on the support $A$ we conclude that the reduced density matrix $\rho_A$ has at most $\binom{N}{k}$ non-zero eigenvalues. Hence, the reduced density matrix $\rho_A(k)$ is a rank $\binom{N}{k}$ matrix and we may posit an upper bound for the entanglement entropy extractable from the projected state as $S_A(k) \le \log_2\{N!/[k!(N-k)!]\} \equiv S_\text{ub}(k)$ if we assume a flat entanglement spectrum.

The eigenvalue problem \eqref{eigen} can be solved analytically by linear algebra methods for small values of $N$, but becomes increasingly cumbersome for large $N$. This difficulty arises because the projected wavefunctions no longer have the convenient structure of the ground state Bethe ansatz, that is, the projected states are excited and the dimensions of the reduced density matrix scale exponentially with particle number. A numerically exact evaluation of the von Neumann entropy is shown in Fig.~\ref{entanglement} for two, four, and six bosons. For the balanced cases we find that more entanglement is present in the projected states as the repulsion strength is increased, with significant entanglement produced above the scale set by $c=1$. In the inset of Fig.~\ref{entanglement} we show that we can extract more entanglement from the balanced projection than from any of the unbalanced cases for $N=4$ and any $c$. This fact is supported, but not rigorously proven for arbitrary $N$, by our calculation in the previous paragraph of the upper bound for the von Neumann entropy: $S_\text{ub}(k\ne N/2)<S_\text{ub}(N/2)$ or, equivalently, $\rank\rho_A(k\ne N/2)<\rank\rho_A(N/2)$. In fact, for all the cases we have considered above the projected states $\chi_{AB}$ are not maximally entangled, that is, $S_A < S_\text{ub}$. Nevertheless, because of the preceding observations and the symmetry $S_A(k) = S_A(N-k)$, we conjecture that $S_{A}(k\ne N/2)<S_{A}(N/2)$ for all $c$.

\begin{figure}[tb]
	\centering
		\includegraphics[width=0.9\linewidth]{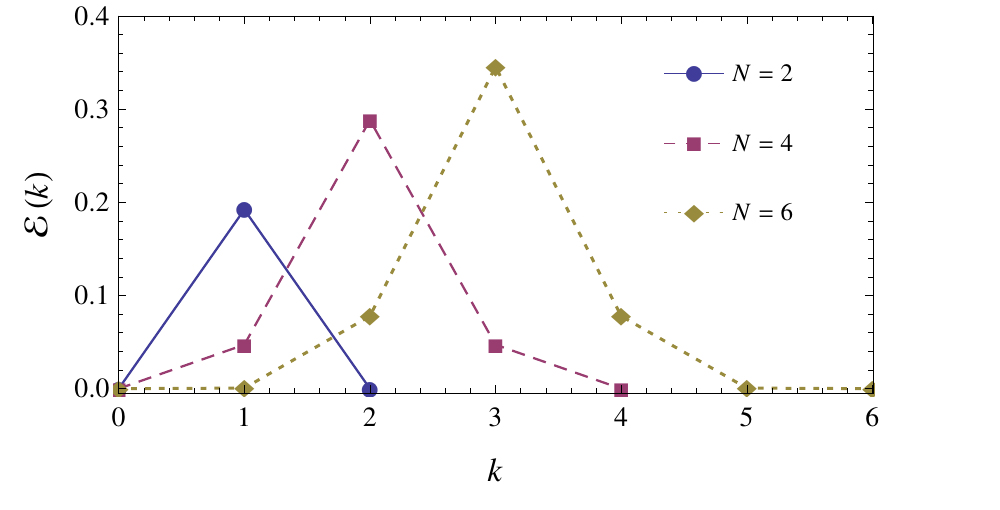}
	\caption{(Color online) In the impenetrable boson limit the weighted entanglement $\mathcal{E}(k)$ is a maximum for the balanced case $k=N/2$. The extractable entanglement $\mathcal{E}_{P\negthinspace P}$ corresponds to $\mathcal{E}(N/2)$. Connecting lines serve to guide the eye.\label{unbalanced}}
\end{figure} 

So far we have argued that the projection probabilities and von Neumann entropies are largest for the balanced projections $k=N/2$ at any given value of the repulsion strength. Thus, according to the definition \eqref{epp}, the projectively extractable pure state entanglement from the Lieb-Liniger ground state is given by the weighted entanglement of the balanced fixed number projections,
\begin{equation}
		\mathcal{E}_{P\negthinspace P} =p(N/2)S_{A}(N/2) = \mathcal{E}_{k=N/2}, \quad \forall c\ge0.
\end{equation}
This statement is trivial for free bosons as we have already shown that the von Neumann entropy vanishes in all possible projections for any even $N$. In the opposite limit of impenetrable bosons this assertion is verified in Fig.~\ref{unbalanced} where we give numerically exact results for all possible cases of up to six bosons. Also apparent in this graph is a slower than linear increase in the extractable entropy $\mathcal{E}_{P\negthinspace P}$ with respect to the boson number $N$.

We also observe that although the probability of successful projection becomes smaller with increasing boson number, the von Neumann entropy $S_A$ increases faster with $N$ so that the extractable entanglement $\mathcal{E}_{P\negthinspace P}$ increases with both repulsion strength and the number of bosons in the ring (Fig.~\ref{eppfig}). Furthermore, for the few boson cases we have analyzed here, this increase is monotonic with respect to both $N$ and $c$. Hence, more entanglement can be extracted from these projections of the Lieb-Liniger gas in the TG limit of impenetrable bosons and we may regard the quantity $\mathcal{E}_{P\negthinspace P}$ as a probe of both quantum correlations and interparticle interactions in the ensemble of fixed number projections of the type we considered here. 

\section{Concluding remarks}\label{conclusions}

\begin{figure}[tb]
	\centering
		\includegraphics[width=0.9\linewidth]{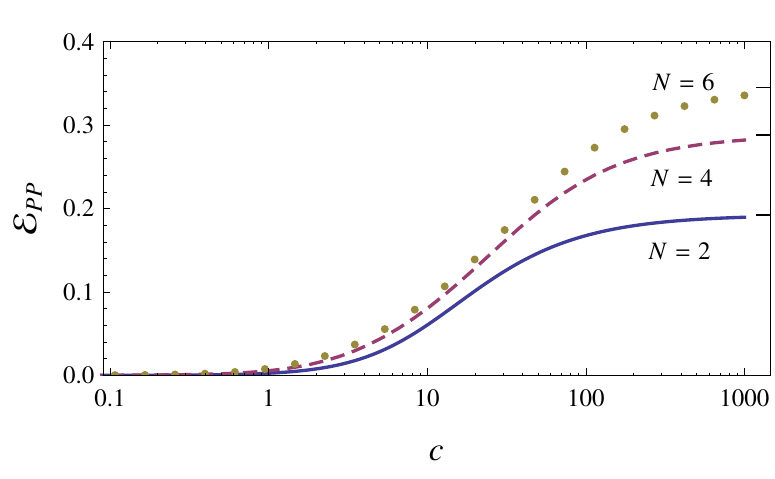}
	\caption{(Color online) The projectively extractable pure state entanglement monotonically increases with repulsion strength, smoothly transitioning from the free and impenetrable boson limits (black ticks).\label{eppfig}}
\end{figure} 

We have used the projectively extractable pure state entanglement $\mathcal{E}_{P\negthinspace P}$ to quantify the entanglement in coarse-grained fixed number projections of the Lieb-Liniger ground state. This entanglement measure quantifies the entanglement present in the set of all projection outcomes by giving the largest von Neumann entropy weighted by the projection probability of a particular measurement result. In our case, this maximum corresponds to the most probable and most entangled projection having an equal number of particles in each half of the ring. We have seen in our numerically exact few particle results indications that the extractable entanglement increases monotonically with the strength of repulsion and with particle number. We have also observed that significant amounts of entanglement can be extracted by this projection procedure only in the strongly repulsive regime $c\gg1$. This increase and subsequent saturation of entanglement with interaction strength $c$ has also been previously observed numerically in a few-particle Lieb-Liniger gas under the framework of single-particle partioning of the ground state \cite{guo}.


As the impenetrable TG limit corresponds to free fermions, our work reveals a fundamental difference in the origin of the entanglement present in bosonic and fermionic systems. When fluctuations in partition occupancies are nullified by appropriate measurements, the entanglement of free-bosonic systems vanish, while for free-fermionic systems they remain non-zero. In the interpolation between the two cases, as is evident from Figs.~\ref{entanglement} and \ref{eppfig}, the free-bosonic behavior seems to last for $c$ nearly up to unity, before a rapid rise in entanglement is seen. A qualitative explanation of the difference found numerically in this paper is the fact that all bosons are exactly in the same state in the $c=0$ limit, so that no uncertainty remains as to their state in each partition once their number in the partition is determined. On the other hand, for free fermions, all states up to the Fermi level will be occupied resulting in uncertainties (entropies) of the state in each partition even when the exact number of fermions in the partition is known. We expect these features to be preserved at larger particle numbers $N$ and ultimately present itself as an essential difference between a Bose-Einstein condensate and a Fermi gas from the perspective of quantum information---a suggestion that is supported by our results and the observation that the Lieb-Liniger model has no density-controlled phase transition at any finite positive density. 

It is important to remark that the projected states we have described here are not eigenstates of the original Lieb-Liniger hamiltonian \eqref{hamil} and will therefore evolve non-trivially in time. The results we present here are therefore only valid immediately after measurement while the spatial partitioning of particles is meaningful; At arbitrary times after measurement the number of particles in each partition will display quantum fluctuations. The time evolution of these projected states and the quantification of the entanglement in them are left as open problems for future work.

\subsection*{Acknowledgments}
F.~P.~and V.~K.~acknowledge financial support by the National Science Foundation through Grant No.~DMS-0905744. Portions of this work was carried out by F.~P.~under the supervision of A.~Hosoya at the Tokyo Institute of Technology and with support from the Foreign Graduate Student Invitation Program. J.~M.-V.~acknowledges financial support from the Spanish Office for Science Grant FIS2009-13483-C02-02 and Fundaci\'on S\'eneca Regi\'on de Murcia Grant 11920/PI/09. S.~B.~acknowledges the support of the UK Engineering and Physical Sciences Research Council, the Royal Society, and the Wolfson Foundation. The authors thank A.~G.~Abanov, Y.~Shikano, and H.~Katsura for helpful conversations.

\subsection*{Appendix}
We restore length units in this Appendix.

We seek an asymptotic approximation to the characteristic function $f(\alpha) \equiv \langle e^{i\alpha\int_0^\ell \psi^\dag(x) \psi(x)\,dx} \rangle$ for the probability distribution $p(k)$ of finding $k$ impenetrable bosons in an arc $\ell$ of a ring of circumference $L$. In the thermodynamic limit $N\to\infty$ with finite particle density $N/L$, the characteristic function is equal to the Fredholm determinant \cite{korepin1}
\begin{equation}\label{freddet}
	f(\alpha) = \det \bigl(\hat{\openone}-(1-e^{i\alpha})\hat{V}\bigr),
\end{equation}
where the linear integral operator $\hat{V}$ acts on the interval $[-q,q]$ with $q = (N-1)\pi/L$ and possesses the kernel
\begin{equation}
	V(\lambda,\mu) = \frac{1}{\pi(\lambda-\mu)} \sin\bigl[ \tfrac{1}{2}(\lambda-\mu)\ell\bigr].
\end{equation}
Let us make a discrete approximation to the integral $\hat{V}[F(\mu)](\lambda)$ by transforming $\lambda \to \lambda_m = (2m-1)\pi/L$ and $\mu \to \mu_n = (2n-1)\pi/L$ to obtain
\begin{equation}
	\int_{-q}^{q} V(\lambda_i,\mu)F(\mu)\,d\mu \approx \sum_{j=1}^N \Gamma_{ij}\, F\biggl[\frac{2\pi}{L}\biggl(j-\frac{N+1}{2}\biggr) \biggr],
\end{equation}
Here, the elements $\Gamma_{ij}$ of the matrix $\bm\Gamma$ are
\begin{equation}
	\Gamma_{ij} = \delta_{ij}\frac{\ell}{L} + (1-\delta_{ij})\frac{\sin[\pi(i-j)\ell/L]}{\pi(i-j)}.
\end{equation}
The characteristic function may therefore be approximated by the $N\times N$ Toeplitz determinant when $N\gg 1$:
\begin{equation}\label{toepdet}
	f(\alpha) \approx \det \bigl(\openone_N-(1-e^{i\alpha}){\bm\Gamma}\bigr).
\end{equation}
The matrix $\bm \Gamma$ is identical to the single-particle correlation matrix of free fermions on an infinite one-dimensional lattice upon making the replacement $\pi\ell/L \to k_Fa$ with $k_F$ the Fermi wavevector and $a$ the lattice spacing \cite{abanov}.

For $\left|{\alpha}\right| < \pi$ the Fisher-Hartwig formulas \cite{basor,deift} may be used to obtain the asymptotic result \cite{[{}][{.\ We note that the correct formula Eq. \eqref{barnes} for the characteristic function has both Barnes $G$-functions squared.}]ovchinnikov,abanov}
\begin{equation}\label{barnes}
	f(\alpha)\sim \frac{e^{i\alpha N \ell/L}\bigl[G\bigl(1+\alpha/(2\pi)\bigr)G\bigl(1-\alpha/(2\pi)\bigr)\bigr]^2}{\bigl[2N\sin(\pi \ell /L)\bigr]^{\alpha^2/(2\pi^2)}},
\end{equation}
where $G(z)$ is the Barnes $G$-function defined functionally through $G(1+z) = \Gamma(z)G(z)$ and $G(1) = 1$, with $\Gamma(z)$ the usual gamma function. Using a small $\alpha\ll\pi$ expansion \cite{abanov} gives
\begin{multline}
	\ln f(\alpha) \sim \frac{i N \ell}{L} \alpha -\frac{\ln[2Ne^{\gamma+1}\sin(\pi\ell/L)]}{2\pi^2}\alpha^2 \\
	+ 0\alpha^3-\frac{\zeta(3)}{(2\pi)^4}\alpha^4, \quad 0<\alpha\ll \pi,
\end{multline} 
where $\gamma$ is Euler's constant. We extract the first two cumulants of the probability distribution $p(k)$ from this expression and make a Gaussian approximation for the case of interest $\ell = L/2$ about the central peak $k = N/2$ to obtain
\begin{equation}
	p(k) \approx \frac{e^{-(k-N/2)^2/(2\sigma^2)}}{\sqrt{2\pi\sigma^2}},
\end{equation}
with variance $\sigma^2 = \pi^{-2} \ln[2Ne^{\gamma+1}]$. Physically, the variance $\sigma^2$ is the fluctuation of particle number in the half-ring $\ell=L/2$ about the mean value $N/2$. The probability of finding exactly $N/2$ impenetrable bosons (or free fermions) in a half-ring is therefore
\begin{equation}
	p(N/2) \sim \sqrt{\frac{\pi}{2\ln(2Ne^{\gamma+1})} }, \quad N\gg 1.
\end{equation}

\end{document}